\documentclass{PoS}

\newcommand{\bra}{\langle}
\newcommand{\ket}{\rangle}
\newcommand{\Tr}{\mbox{Tr}\,}
\renewcommand{\Re}{\mbox{Re}\,}
\renewcommand{\Im}{\mbox{Im}\,}
\newcommand{\be}{\begin{equation}}
\newcommand{\ee}{\end{equation}}
\newcommand{\bea}{\begin{eqnarray}}
\newcommand{\eea}{\end{eqnarray}}
\newcommand{\nn}{\nonumber}
\newcommand{\xv}{{\mathbf x}}
\newcommand{\cP}{{\cal P}}
\newcommand{\eps}{\epsilon}
\newcommand{\id}{1\!\!\!1}

\title{Stochastic quantization at nonzero chemical potential}

\ShortTitle{Stochastic quantization at nonzero chemical potential}

\author{\speaker{Gert Aarts}\\
        Department of Physics, Swansea University, Swansea, United Kingdom\\
        E-mail: \email{g.aarts@swan.ac.uk}}

\author{Ion-Olimpiu Stamatescu\\
        Institut f\"ur Theoretische Physik, Universit\"at Heidelberg and 
        FEST, Heidelberg, Germany\\
        E-mail: \email{I.O.Stamatescu@thphys.uni-heidelberg.de}}

\abstract{ 
 Lattice QCD at finite chemical potential is difficult due to the 
sign problem. We use stochastic quantization and complex Langevin 
dynamics to study this issue. First results for QCD in the hopping 
expansion are encouraging. U(1) and SU(3) one link models are used to gain 
further insight into why the method appears to be successful.
 }

\FullConference{The XXVI International Symposium on Lattice Field Theory\\
		 July 14-19 2008\\
		 Williamsburg, Virginia, USA}

\begin{document}

\section{Introduction}

QCD at finite chemical potential is difficult because the complex fermion 
determinant prohibits lattice techniques based on direct importance 
sampling, while reweighting methods suffer from sign and overlap problems. 
Stochastic quantization \cite{Parisi:1980ys} is an alternative 
nonperturbative method which in principle can deal with complex actions, 
via the use of complex Langevin dynamics 
\cite{Parisi:1984cs,Klauder:1985a,Klauder:1985b,Gausterer:1986gk}. 
However, since proposed in the 80's, progress has been hindered by 
numerical instabilities and uncertainty about or lack of convergence (see 
e.g.\ Refs.\ \cite{Karsch:1985cb,Ambjorn:1986fz} for relevant work). 
Recently it was shown in the context of nonequilibrium (Minkowski) quantum 
field dynamics that some of these problems could be alleviated by the use 
of more refined Langevin algorithms 
\cite{Berges:2005yt,Berges:2006xc,Berges:2007nr}. These studies motivated 
us to consider euclidean systems at finite chemical potential, either 
derived from or with a structure similar to QCD at finite density. The 
first results are very promising and can be found in Ref.\ 
\cite{Aarts:2008rr}.

\section{Three models with a sign problem and complex Langevin dynamics}

We consider three models inspired by or derived from QCD. The partition 
functions have the familiar form
 \be
 Z = \int DU\, e^{-S_B} \det M,
\;\;\;\;\;\;\;\;\; 
\mbox{with}
\;\;\;\;\;\;\;\;\; 
\det M(\mu) = [\det M(-\mu)]^*.
 \ee
 Here $S_B$ is the real bosonic action, depending on the gauge links $U$,  
and $\det M$ is the complex fermion determinant, with $\mu$ the chemical 
potential. 

\underline{QCD in the hopping expansion}: The (Wilson) fermion matrix 
reads
\be
 M = 1 - \kappa\sum_{i=1}^3 \mbox{space}
-\kappa \left( e^\mu \Gamma_{+4} U_{x,4}T_{4} +
e^{-\mu}\Gamma_{-4} U_{x,4}^{-1} T_{-4} \right).
\ee
 In the hopping expansion at nonzero $\mu$, $\kappa$ is taken to $0$ but 
terms with $\kappa e^{\pm\mu}$ are preserved. As a result, only the 
temporal links survive and the determinant is local,
\bea
\nn
\det M &\approx& \det \left[ 1
-\kappa \left( e^\mu \Gamma_{+4} U_{x,4}T_{4} +
e^{-\mu}\Gamma_{-4} U_{x,4}^{-1} T_{-4} \right) \right]
\\
&=& \prod_{\xv}
\det\left( 1 + h e^{\mu/T} \cP_{\xv} \right)^2
\det\left( 1 + h e^{-\mu/T} \cP_{\xv}^{-1} \right)^2.
\eea
 Here $h = (2\kappa)^{N_\tau}$ and $\cP_\xv^{(-1)}$ are the (conjugate) 
Polyakov loops. 
 The gauge action is the standard Wilson action, $S_B=-\beta \sum_P 
(\frac{1}{6}[\Tr U_P + \Tr U_P^{-1}] -1)$ and is preserved completely.
 
In order to investigate the algorithm in detail, we have also studied two 
one link models, where exact results are available. 

\underline{SU(3) one link model}: The bosonic action is 
$S_B=-\frac{\beta}{6} (\Tr U + \Tr U^{-1})$, with $U\in$ SU(3). The 
determinant has the product form as above,
\be
 \det M = \det\left( 1 + \kappa e^{\mu} U\right) \left( 1 + \kappa 
e^{-\mu} 
U^{-1}\right).
\ee
 Exact results follow from integration over the reduced Haar measure.

\underline{U(1) one link model}: The bosonic action is 
$S_B=-\frac{\beta}{2} (U + U^{-1})$ and the determinant is $\det 
M=1+\frac{\kappa}{2} \left(e^\mu U+e^{-\mu}U^{-1}\right)$. When the link 
is written as $U=e^{ix}$, the partition function is a one-dimensional 
integral,
\be
 Z =  \int_{-\pi}^\pi\frac{dx}{2\pi}\, e^{\beta\cos x}
\left[   1+\kappa \cos(x-i\mu) \right],
\ee
 and exact expressions are available in terms of Bessel functions.

All models behave in the same way as QCD under the substitution $\mu\to 
-\mu$. Observables investigated are the (conjugate) Polyakov loops, the 
density and the phase of the determinant.

 Since the determinants are complex, standard lattice methods cannot be 
used. We employ complex Langevin dynamics instead. 
The Langevin update [for SU(3)] is
\be
 \label{eqlang}
 U(\theta+\eps)  = R(\theta)\, U(\theta),
\;\;\;\;\;\;\;\;
\;\;\;\;\;\;\;\;
R = \exp \left[ i\lambda_a\left( \eps K_a +\sqrt \eps \eta_a\right)
\right],
\ee
 where $\theta$ is the Langevin time, $\eps$ is the stepsize, $\lambda_a$ 
($a=1,\ldots,8$) are 
the Gell-Mann matrices, the drift term reads
\be
 K_a =  -D_a S_{\rm eff},
\;\;\;\;\;\;\;\;\;\;\;\;
 S_{\rm eff} = S_B+S_F,
 \;\;\;\;\;\;\;\;\;\;\;\;
 S_F = - \ln\det M,
\ee
with $D_a f(U)= \partial_\alpha f(e^{i\alpha \lambda_a} U)|_{\alpha=0}$,
and the noise 
satisfies
\be
 \bra\eta_a\ket=0,
\;\;\;\;\;\;\;\;\;\;\;\;\;\;\;\;
 \bra\eta_a\eta_b\ket=2\delta_{ab},
\ee
 suppressing euclidean spacetime indices.
 Since the action and as a result the drift term are complex, $R^\dagger 
R\neq \id$, although $\det R=1$ still holds. Therefore the complex 
Langevin dynamics takes place in SL(3, $\mathbb{C}$) and not in SU(3). We 
come back to this below. Related to this, we note that after 
complexification observables should be defined in terms of $U$ and 
$U^{-1}$, but not of $U^\dagger\neq U^{-1}$.

\section{Results}

The complex Langevin equation (\ref{eqlang}) was solved numerically, with 
a stepsize $\eps=2-5\times 10^{-5}$. In the one link models, we have not 
observed any instability or runaway solution. In the field theory, 
runaways have been eliminated by being careful with numerical precision 
and roundoff errors, and employing a dynamical stepsize. Here we show some 
results for illustration; a more extensive discussion can be found in 
Ref.\ \cite{Aarts:2008rr}.

\begin{figure}
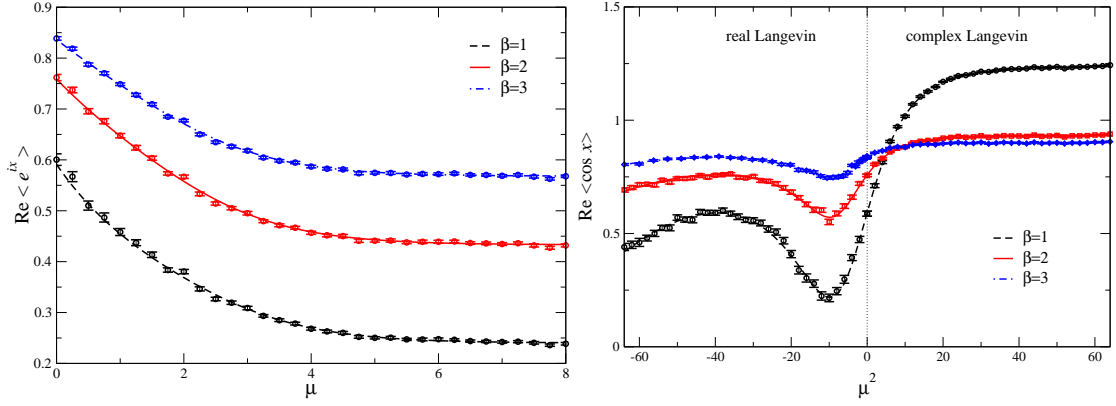

\begin{center}
\includegraphics[height=5.3cm]{plot_full_k0.25_poly.eps}
\includegraphics[height=5.3cm]{plot_full_b1_k0.5_mu2.eps}
 \caption{Real part of the Polyakov loop $\bra e^{ix}\ket$ as a function 
of $\mu$ (left) and the plaquette $\bra\cos x\ket$as a function of $\mu^2$ 
(right) for three values of $\beta$ at fixed $\kappa=1/2$ in the U(1) one 
link model. The lines are the analytical results, the symbols are obtained 
with Langevin dynamics.
 }
\label{figu1}
\end{center}
\end{figure}

The Polyakov loop $\Re\bra e^{ix}\ket$ in the U(1) model is shown in Fig.\ 
\ref{figu1} (left). The data points come from the Langevin dynamics, the 
lines are the exact solutions. Excellent agreement is observed. At 
imaginary chemical potential, $\mu=i\mu_I$, the determinant is real and 
there is no need to complexify the Langevin dynamics. The smooth 
connection between the results obtained with complex Langevin (when 
$\mu^2>0$) and real Langevin (when $\mu^2<0$) is shown in Fig.\ 
\ref{figu1} (right) for the plaquette $\bra \cos x\ket$. In particular, 
statistical errors are comparable. We note here that also in the SU(3) one 
link model, excellent agreement between exact and numerical results can be 
seen \cite{Aarts:2008rr}.

\begin{figure}
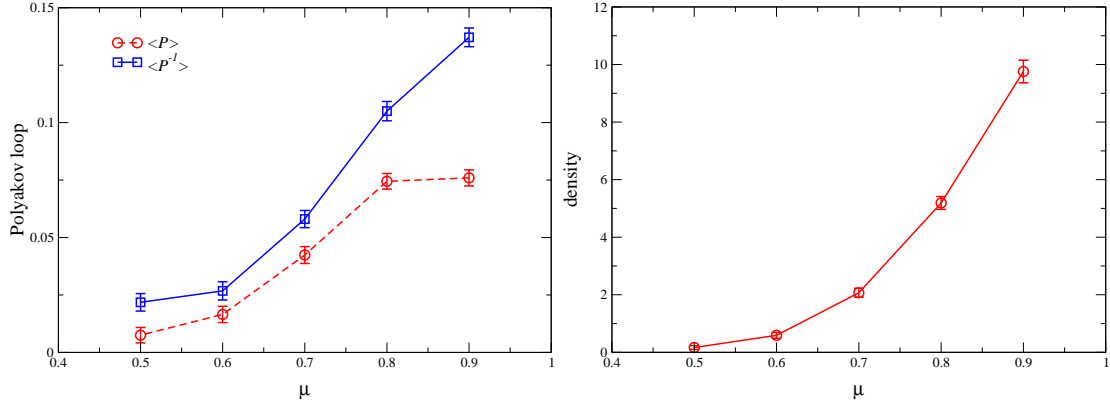

\begin{center}
\includegraphics[height=5.3cm]{plot_fsu3_rePPcc_Nt4_ch2.eps}
\includegraphics[height=5.3cm]{plot_fsu3_redensity_Nt4_ch2.eps}
\end{center}
 \caption{
Real part of the Polyakov loop $\bra P\ket$ and the conjugate
Polyakov loop  $\bra P^{-1}\ket$ (left) and the density $\bra 
n\ket$ (right) as a function of $\mu$ at $\beta=5.6$, $\kappa=0.12$ on
a $4^4$ lattice, with $N_f=3$ flavours.
 }
\label{figfsu3}
\end{figure}

First results for the QCD in the hopping expansion are shown in Fig.\ 
\ref{figfsu3} for the (conjugate) Polyakov loops $P^{(-1)} = 
\frac{1}{3}\Tr \cP^{(-1)}$ (left) and the density $\bra n\ket= 
T\partial\ln Z/\partial\mu$, with $T=1/N_\tau$ (right). These results are 
obtained on a $4^4$ lattice at $\beta=5.6$. We observe that the 
(conjugate) Polyakov loops increase from a small value at $\mu=0.5$ to a 
value clearly different from zero at the larger $\mu$ values. Similarly, 
the density increases substantially with chemical potential. We interpret 
this as indications for a transition from a low-density ``confining'' 
phase to a high-density ``deconfining'' phase.

\begin{figure}
\begin{center}
\includegraphics[height=5.3cm]{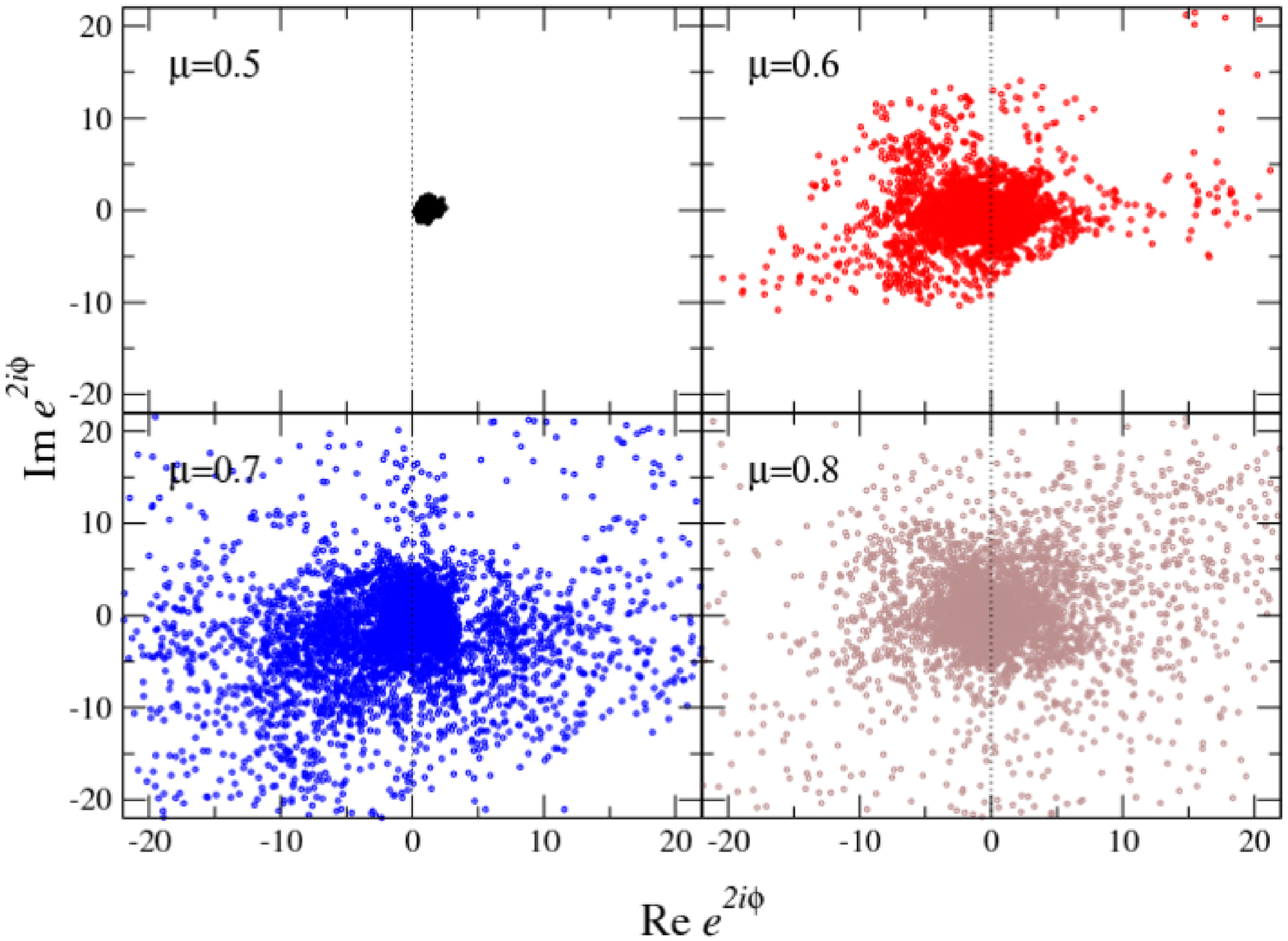}
\includegraphics[height=5.3cm]{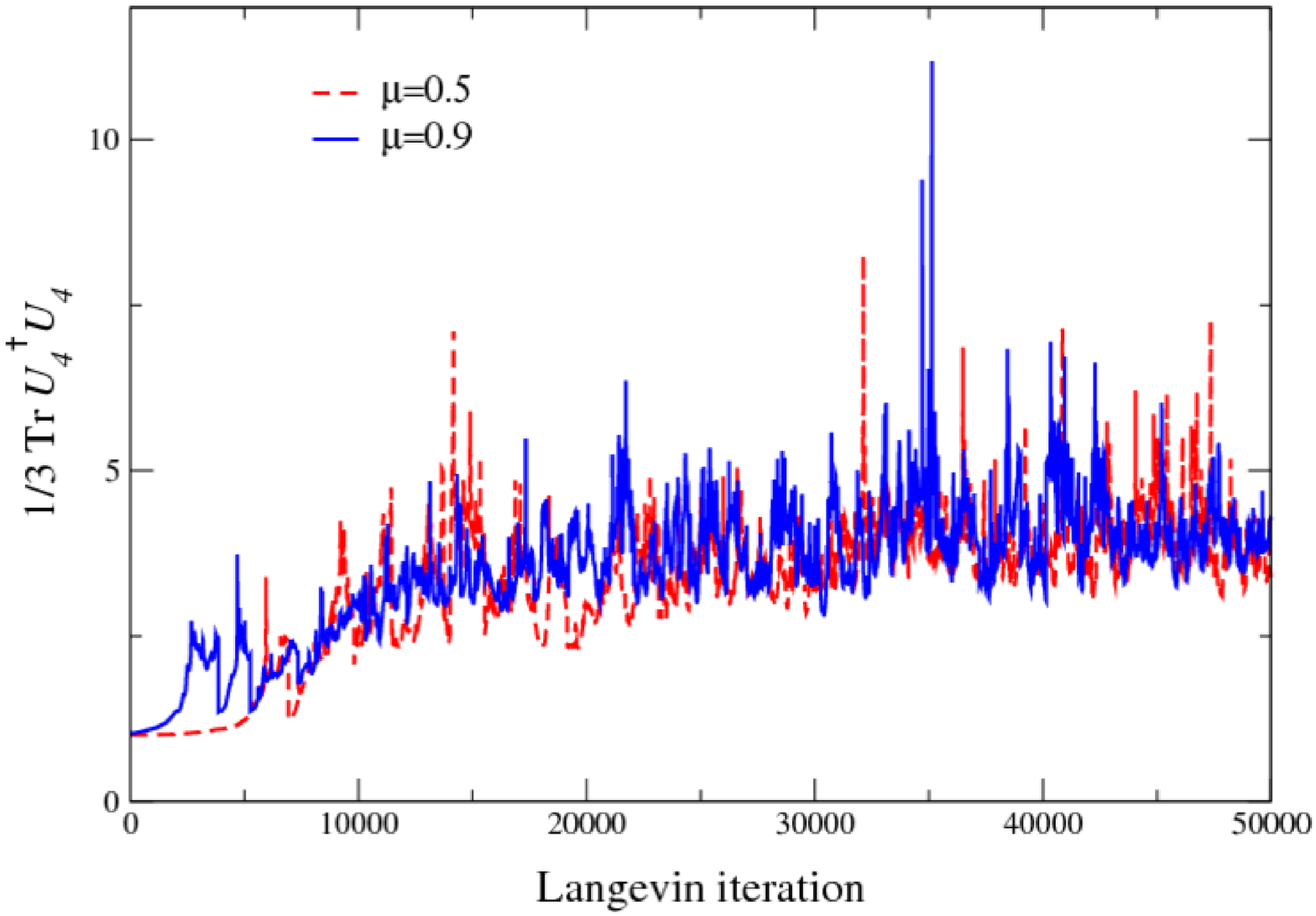}
\end{center}
 \caption{
Left: Scatter plot of $e^{2i\phi} = \det M(\mu)/\det
M(-\mu)$ during the Langevin evolution for various values of $\mu$ at
$\beta=5.6$, $\kappa=0.12$ on a $4^4$ lattice. Right:
Deviation from SU(3), $\Tr U_4^\dagger U_4/3$ during the Langevin
evolution, for $\mu=0.5$ and $0.9$.
 }
\label{figscat}
\end{figure}

The sign problem in these models can be studied by writing the determinant 
as $\det M(\mu) = [\det M(-\mu)]^* = |\det M(\mu)| e^{i\phi}$ and 
considering the average phase factor
\be
 \bra e^{2i\phi}\ket = \left\bra \frac{\det M(\mu)}{\det 
M(-\mu)}\right\ket. 
\ee 
 Scatter plots of $e^{2i\phi}$ during the Langevin evolution in QCD in the 
hopping expansion are shown in Fig.\ \ref{figscat} (left). At zero 
chemical potential $\Re e^{2i\phi}=1$, $\Im e^{2i\phi}=0$. At nonzero 
$\mu$, we observe that phase fluctuations suddenly increase enormously. 
This behaviour is not unexpected when the sign problem is severe. For 
instance, the variance of the phase, $\bra\phi^2\ket-\bra\phi\ket^2$, is 
large and proportional to the four volume $N_\sigma^3N_\tau$. We 
emphasize, however, that the observables (Polyalov loop, density) are 
under control, with a reasonable numerical error. This suggests that phase 
fluctuations and the sign problem may not be a problem for this approach.

Because of the complexification, the dynamics no longer takes place in 
SU(3) but in SL(3, $\mathbb{C}$) instead. When the links are written in 
terms of gauge potentials, $U=e^{i\lambda_aA_a/2}$, this implies that the  
$A_a$'s are now complex. A measure of how much the dynamics deviates from 
SU(3) can be given by considering $\frac{1}{N}\Tr U^\dagger U$ which $=1$ 
if $U\in$ SU($N$) and $\geq 1$ if $U\in$ SL($N$, $\mathbb{C}$). Note that 
this observable is not analytic in $U$; it therefore does not correspond 
to an observable in the original gauge theory before complexification. In 
Fig.\ \ref{figscat} (right) we show the Langevin time dependence of 
$\frac{1}{3}\Tr U_4^\dagger U_4$ on the $4^4$ lattice for $\mu=0.5$ and 
$0.9$. After the thermalization stage, we observe a distinct deviation 
from 1, as expected. What is important is that the observable remains 
bounded during the evolution and does not run away to infinity.

\begin{figure}
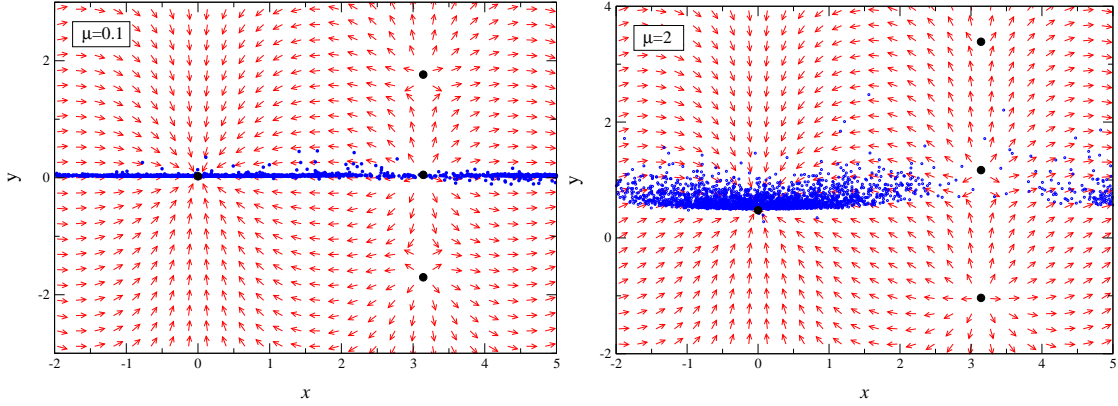

\begin{center}
\includegraphics[height=5.3cm]{plot_full_b1_k0.25_m0.1_scat.eps}
\includegraphics[height=5.3cm]{plot_full_b1_k0.25_m2_scat.eps}
\end{center}
 \caption{Classical flow diagram in the U(1) one link model. The 
horizontal $x$ (vertical $y$) axis corresponds to the real (imaginary) 
part of the 
gauge potential. The big dots indicate the fixed points at
$x=0$ and $\pi$. The small circles indicate a trajectory during the
Langevin evolution. Note the periodicity $x\to x+2\pi$. Parameters are
$\beta=1$, $\kappa=1/2$, $\mu=0.1$ (left) and $\mu=2$ (right).
 }
\label{figflow}
\end{figure}

Some insight into why stochastic quantization works at finite chemical 
potential can be obtained from the simple U(1) one link model. Consider 
the link $U=e^{ix}$ after complexification, with $x\to x+iy$. The Langevin 
equations are $\dot x = K_x+\eta$, $\dot y = K_y$, where $K_{x,y}$ are 
classical forces and the dot indicates a $\theta$ derivative. 
Classical flow diagrams in the $x$--$y$ plane are shown in Fig.\ 
\ref{figflow}. We find a stable fixed point at $x=0$ and unstable fixed 
points at $x=\pi$. The important observation is that this structure is 
independent of $\mu$. The small (blue) dots indicate a trajectory during 
the Langevin evolution. As is clearly visible, in the vertical (unbounded) 
direction the dynamics is attracted to the stable fixed point and remains 
bounded.

 These flow diagrams are also useful to illustrate how the method is 
distinct from other approaches based on employing configurations obtained 
at zero chemical potential (reweighting, Taylor expansion). In the 
language of this simple model, in those approaches configurations are 
generated without an imaginary component of the gauge potential (i.e.\ 
$y=0$), and these are subsequently used to probe the system at finite 
chemical potential. In contrast, in this approach configurations have 
nonzero imaginary parts of the gauge potential ($y\neq 0$) and change 
therefore in an essential way when $\mu$ departs from zero, as indicated
by comparing the two plots in Fig.\ \ref{figflow}.

\begin{figure}
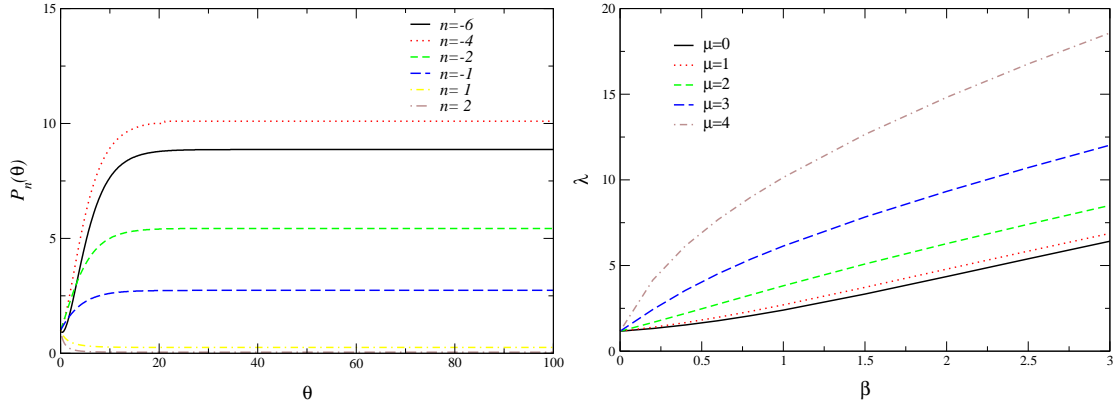

\begin{center}
\includegraphics[height=5.3cm]{fp_b1_k0.5_mu3.eps}
\includegraphics[height=5.3cm]{fokker_eigenvalues_k0.5_mu0-4.eps}
\end{center} 
 \caption{Left: Langevin evolution of the modes $P_n(\theta)$ of the 
complex Fokker-Planck distribution in the U(1) one link model ($\beta=1$, 
$\kappa=1/2$, $\mu=3$). 
Right: Smallest nonzero eigenvalues of the complex Fokker-Planck operator 
as a function 
of $\beta$ for various values of $\mu$ at $\kappa=1/2$.
 } 
\label{figfp} 
\end{figure}

A second indication for the success of this approach in the U(1) 
model comes from an analysis of the complex Fokker-Planck equation for the 
$\theta$ dependent distribution $P(x,\theta)$,
\be
\frac{\partial P(x,\theta)}{\partial \theta} =
\frac{\partial}{\partial x}\left(
\frac{\partial}{\partial x} + \frac{\partial S}{\partial x}
\right) P(x,\theta).
\ee
 We have solved this equation for the modes $P_n(\theta)=\int_{-\pi}^\pi 
e^{inx}P(x,\theta)/2\pi$ and a typical result is shown in 
Fig.\ \ref{figfp} (left). A rapid convergence to the correct distribution 
$P(x)\sim e^{-S(x)}$ is observed.  
 This behaviour can be understood from the eigenvalues of the complex 
Fokker-Planck operator. It follows from the symmetries of the model under 
$\mu\to -\mu$, $x\to -x$ that the eigenvalues are real. We found 
numerically that the nonzero eigenvalues are positive definite, see Fig.\ 
\ref{figfp} (right). Although these results are not sufficient to prove 
convergence of the complexified dynamics analytically, it supports the 
stochastic results presented above.

\section{Conclusion}

We have considered complex Langevin dynamics to study theories with a 
complex action due to a chemical potential. 
 In the U(1) and SU(3) one link models the agreement between exact and 
stochastic results is excellent. Moreover, in these simple models it is 
possible to gain insight into why the method works, using classical flow 
diagrams and an analysis of the eigenvalues of the complex Fokker-Planck 
operator.
 First results in QCD in the hopping expansion are very encouraging. Even 
though the phase of the determinant is fluctuating wildly, observables 
such as the Polyakov loop and the density can be measured with reasonable 
errors.
 Given these findings and the experience we have obtained so far, we are 
led to believe that stochastic quantization might be insensitive to the 
sign problem. Unpublished results \cite{Aarts:2008} support these 
conclusions.

\acknowledgments

We thank Erhard Seiler who collaborated during part of
this work and contributed many important insights. 
 G.A.\ thanks Simon Hands, Biagio Lucini and Asad Naqvi for
discussion.
 We thank the Kavli Institute for Theoretical Physics in Santa Barbara, 
the Yukawa Institute for Theoretical Physics in Kyoto, and the 
Max-Plank-Institute (Werner Heisenberg Institute) in Munich for 
hospitality during the time in which this work was carried 
out.
 G.A.\ is supported by STFC.

\end{document}